\documentclass[letter]{jpsj2} 

\def\runauthor{S. Watanabe, M. Imada and K. Miyake}

\title{%
Superconductivity Emerging near Quantum Critical Point of Valence Transition
}

\author{%
Shinji Watanabe\thanks{E-mail address: swata@issp.u-tokyo.ac.jp}$^1$, 
Masatoshi Imada$^{1,2}$ 
and Kazumasa Miyake$^3$
}
 
\inst{%
Institute for Solid State Physics, 
University of Tokyo, 5-1-5 Kashiwanoha, Kashiwa, Chiba 277-8581, Japan$^1$
\\
PRESTO, Japan Science and Technology Agency, 4-1-8 Honcho, Kawaguchi, Saitama 332-0012, Japan$^2$
\\
Division of Materials Physics, Department of Materials Engineering Science, Graduate School of
Engineering Science, Osaka University, Toyonaka, Osaka 560-8531, Japan$^3$
}

\recdate{\today}

\abst{%
The nature of the quantum valence transition is studied 
in the one-dimensional periodic Anderson model with Coulomb repulsion 
between f and conduction electrons by the density-matrix renormalization group method. 
It is found that the first-order valence transition emerges 
with the quantum critical point and
the crossover from the Kondo to the mixed-valence states is strongly stabilized 
by quantum fluctuation and electron correlation. 
It is found that the superconducting correlation is developed in the Kondo regime 
near the sharp valence increase. 
The origin of the superconductivity is ascribed to the development of 
the coherent motion of electrons with enhanced valence fluctuation, which results 
in the enhancement of the charge velocity, but not of the charge compressibility. 
Statements on the valence transition 
in connection with Ce metal and Ce compounds are given. 
}

\kword{%
$\rm CeCu_2Ge_2$, 
valence transition, quantum critical point, 
valence fluctuation, superconductivity, DMRG, periodic Anderson model 
}

\begin{document}
\maketitle

In $\rm CeCu_2Ge_2$~\cite{Jaccard}, the superconducting transition temperature 
$T_{\rm SC}$ has a maximum in the pressure regime 
far from the antiferromagnetic (AF) quantum critical point (QCP), 
where the coefficient $A$ in the resistivity $\rho=\rho_{0}+AT^n$ drops 
rapidly for $n \sim 2$. 
Since $A$ is proportional to the effective mass~\cite{KM} 
$(m^{*}/m)^2$ which is related to the f-electron number as 
$m^{*}/m=(1-n_{\rm f}/2)/(1-n_{\rm f})$~\cite{Rice}, 
this implies that 
$T_{\rm SC}$ has a maximum around the sharp valence change of Ce. 
Similar behavior has been observed in the isostructural compound 
$\rm CeCu_2Si_2$~\cite{Holmes} and in $\rm CeCu_2(Si_{1-x}Ge_{x})_2$~\cite{Steglich}. 
Recently, an NMR measurement has revealed that 
$T_{\rm SC}$ increases under hydrostatic pressure in $\rm CeIrIn_5$ 
although AF spin fluctuation is suppressed~\cite{Kawasaki}, 
suggesting the different mechanism for the superconductivity 
from ordinary AF spin fluctuation.

As known as the $\gamma$-$\alpha$ transition~\cite{Ce}, Ce metal shows the first-order 
valence transition in its temperature-pressure phase diagram. 
The valence change of Ce is due to 
the 4f level located near the Fermi level, which can hybridize easily with 
the conduction band by applying pressure. 
At the critical point, the valence susceptibility diverges 
as diverging density fluctuation in the liquid-gas transition. 
When the critical temperature is suppressed by controlling the chemical substitution 
and applying pressure, and enters the Fermi-degeneracy regime, 
diverging valence fluctuation is considered to be coupled 
with the Fermi-surface instability. 
This multiple instability in the quantum-degeneracy regime 
seems to be a key mechanism for understanding the instabilities observed 
in the above Ce compounds. 
However, theoretical understanding of the mechanism has not been fully achieved 
although 
a pioneering work on the valence-fluctuation-mediated superconductivity~\cite{OM} 
and later discussions~\cite{Lonzarich} have been made. 

In this Letter, we clarify the nature of the first-order valence transition 
as well as the electronic states near the quantum critical point.  
We show that the superconducting correlation becomes developed 
in the Kondo regime near the sharp valence crossover on the basis of 
the density-matrix renormalization group (DMRG)~\cite{white} calculation 
for the one-dimensional periodic Anderson model
\begin{eqnarray}
H&=& -t \sum_{i\sigma}
(c_{i\sigma}^{\dagger}c_{i+1\sigma}+c_{i+1\sigma}^{\dagger}c_{i\sigma})
+\varepsilon_{\rm f}\sum_{i\sigma}n^{\rm f}_{i\sigma}
\nonumber \\
&+&V\sum_{i\sigma}\left(
f_{i\sigma}^{\dagger}c_{i\sigma}+c_{i\sigma}^{\dagger}f_{i\sigma}
\right)
+U\sum_{i=1}^{N}n_{i\uparrow}^{\rm f}n_{i\downarrow}^{\rm f}
\nonumber
\\
&+&U_{\rm fc}\sum_{i=1}^{N}n_{i}^{\rm f}n_{i}^{\rm c}. 
\label{eq:PAM}
\end{eqnarray}
Here, the notation is standard and 
the number operator is defined by 
$n^{\rm a}_{i\sigma}=a_{i\sigma}^{\dagger}a_{i\sigma}$ and 
$n^{\rm a}_{i}=n^{\rm a}_{i\uparrow}+n^{\rm a}_{i\downarrow}$ for ${\rm a}={\rm f}$ and c.  
The last term is the Coulomb repulsion between f and conduction electrons, 
which is important in causing the valence transition~\cite{HF1,OM,OMV}.

The filling $n$ is defined by $n=(n_{\rm f}+n_{\rm c})/2$ with
$
n_{\rm a}=\sum_{i=1}^{N}\langle n^{\rm a}_{i} \rangle/N
$
for $\rm a=f$ and c 
with $N$ being the number of lattice sites. 
We note that half filling is realized when $n=1$ in this definition.
In this Letter, we consider the case $n=7/8$, $t=1$, $V=0.1$ and $U=100$. 
In the DMRG calculation, 
we have kept the number of states up to 1500 
and the system sizes up to $N=80$ with the open-boundary condition.

\begin{figure}[t]
\begin{center}
\includegraphics[width=8cm]{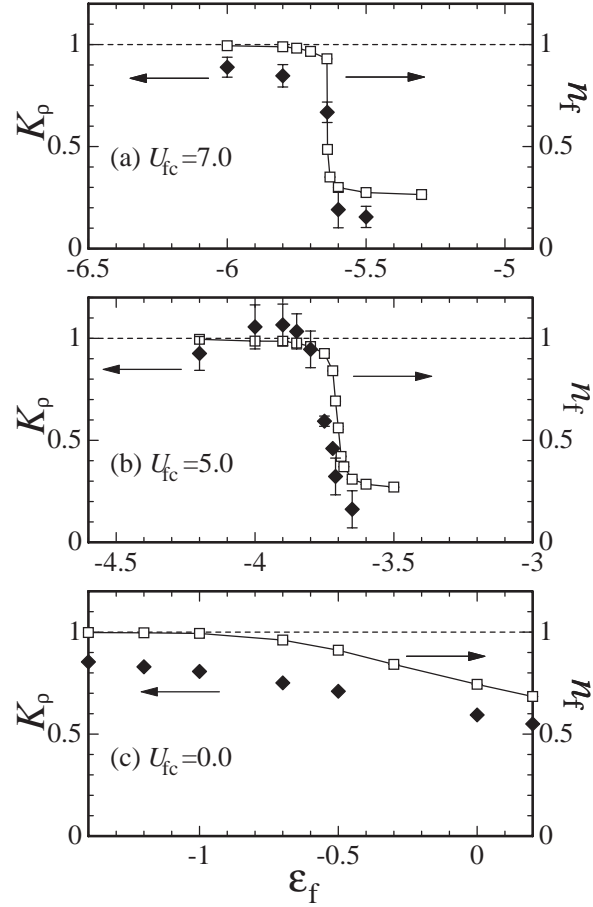}
\end{center}
\caption{$n_{\rm f}$ vs $\varepsilon_{\rm f}$ (open square) 
and $K_{\rho}$ vs $\varepsilon_{\rm f}$ (solid diamond) 
for $t=1$, $V=0.1$ and $U=100$ 
at $n=7/8$ for (a) $U_{\rm fc}=7.0$, (b) 5.0 and (c) 0.0. 
}
\label{fig:nf_Ef}
\end{figure}

Figure~\ref{fig:nf_Ef}(a) shows the f-electron-number $n_{\rm f}$ vs 
the f-level $\varepsilon_{\rm f}$ for $U_{\rm fc}=7.0$, 
which has been 
obtained by a linear extrapolation to $N\to\infty$ using the data for 
$N=24,32,40,48,56,64,72$ and $80$. 
We see that a jump in $n_{\rm f}$ appears at $\varepsilon_{\rm f}=-5.6395$, 
which indicates the first-order quantum phase transition, 
since a jump in $n_{\rm f}$ results in the level crossing of the ground states 
by the relation 
$n_{\rm f}=\partial \langle H \rangle/\partial \varepsilon_{\rm f}$. 
The first-order valence transition (FOVT) 
between the Kondo state with $n_{\rm f}\simeq 1$ and the mixed valence (MV) state 
with 
$n_{\rm f}\simeq1/4$ ($=2(1-n)$ for 1/2 $\le n \le 1$) 
is caused by $U_{\rm fc}$, 
since a large $U_{\rm fc}$ forces the electrons to pour into either the f level or the conduction band.

When $U_{\rm fc}$ decreases, the magnitude of the jump in $n_{\rm f}$ decreases. 
We see that the jump is invisible at $U_{\rm fc}=5.0$ and the sharp crossover 
from the Kondo to MV states appears in Fig.~\ref{fig:nf_Ef}(b). 
These results imply that the QCP at which both the jump disappears  
and the valence susceptibility 
$\chi_{\rm f}\equiv -\partial^{2} \langle H \rangle/\partial \varepsilon_{\rm f}^{2}
=-\partial n_{\rm f}/\partial \varepsilon_{\rm f}$ diverges, 
exists between $U_{\rm fc}=5.0$ and $7.0$. 
As $U_{\rm fc}$ is set to be smaller than 5.0, the change in $n_{\rm f}$ becomes smoother 
and finally a gradual crossover appears at $U_{\rm fc}=0$ as seen in Fig.~\ref{fig:nf_Ef}(c).

To determine the location of the QCP, we have calculated 
the jump of the f-electron number, 
$\Delta n_{\rm f}\equiv n_{\rm f}^{\rm K}-n_{\rm f}^{\rm MV}$, 
at the FOVT point as a function of $\varepsilon_{\rm f}$ 
for several $U_{\rm fc}$ values.   
Here, the superscripts K and MV specify the Kondo and mixed-valence states, respectively. 
Then, the QCP is estimated at 
$(\varepsilon_{\rm f}^{\rm QCP}, U_{\rm fc}^{\rm QCP})\simeq (-4.5, 5.9)$ 
at which the jump disappears. 
In the slave-boson mean-field (MF) theory~\cite{Watanabe} 
the QCP is identified at 
$(\varepsilon_{\rm f \ MF}^{\rm QCP}, U_{\rm fc \ MF}^{\rm QCP})=(0.19, 0.98)$. 
The critical nature of the valence transition will be discussed in detail 
in a separate paper~\cite{Watanabe}.

\begin{figure}[t]
\begin{center}
\includegraphics[width=8cm]{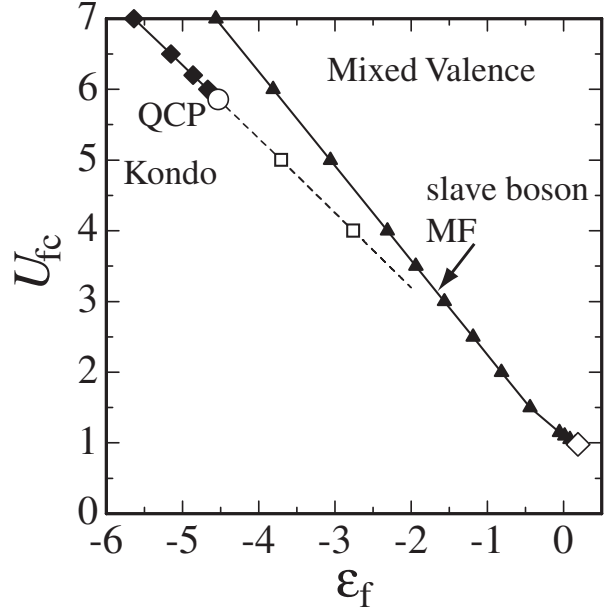}
\end{center}
\caption{
Ground-state phase diagram in $U_{\rm fc}$-$\varepsilon_{\rm f}$ plane 
for $t=1$, $V=0.1$, $U=100$ at $n=7/8$. 
The first-order valence transition line (solid line with solid diamonds) 
with the quantum critical point (open circle) 
is determined by the DMRG. 
The dashed line with open squares represents 
the crossover determined by the point where 
$\chi_{\rm f}$ has its maximum value.
The first-order valence transition line (solid line with solid triangles) 
with the QCP (open diamond) is determined by the slave-boson mean-field theory. 
}
\label{fig:PD}
\end{figure}

We summarize the ground-state phase diagram in the 
$U_{\rm fc}$-$\varepsilon_{\rm f}$ plane in Fig.~\ref{fig:PD}. 
The solid line with solid diamonds represents the FOVT line 
determined by the DMRG, 
and the dashed line with open squares represents the point at which $\chi_{\rm f}$ 
has a maximum, which indicates the crossover with enhanced valence fluctuation.
Here, we note that 
the slope of the FOVT line is expressed as the ratio of the jump in 
$n_{\rm f}$ and $\langle n_{i}^{\rm f}n_{i}^{\rm c}\rangle$ 
at the transition by the `Claudius-Clapeyron relation'~\cite{Thermo}, 
$\delta U_{\rm fc}/\delta \varepsilon_{\rm f}=-(n_{\rm f}^{\rm K}-n_{\rm f}^{\rm MV})
/(C_{\rm fc}^{\rm K}-C_{\rm fc}^{\rm MV})$, 
where $C_{\rm fc}$ is defined as 
$C_{\rm fc} \equiv \sum_{i=1}^{N}\langle n_{i}^{\rm f}n_{i}^{\rm c}\rangle/N$. 
For $U_{\rm fc} \gg U_{\rm fc}^{\rm QCP}$, 
$\delta U_{\rm fc}/\delta \varepsilon_{\rm f}$ is shown to be 
$-1/n_{\rm c}^{\rm K}$ in the MF theory~\cite{Watanabe}, which 
coincides with the equation of the energy valance 
of the one-body f-electron energy at the FOVT; 
$\varepsilon_{\rm f}+U_{\rm fc}n_{\rm c}^{\rm K} \sim \mu$, with 
$\mu$ being the chemical potential~\cite{OM}. 
However, it should be stressed that 
the correct slope for $U_{\rm fc} \gg U_{\rm fc}^{\rm QCP}$ 
is given as $\delta U_{\rm fc}/\delta \varepsilon_{\rm f}=-1$ 
by inserting the atomic-limit values of 
$n^{\rm K}_{\rm f}=1$, $n^{\rm MV}_{\rm f}=1/4$, 
$C^{\rm K}_{\rm fc}=3/4$ and $C^{\rm MV}_{\rm fc}=0$ into the thermodynamic relation. 
This discrepancy is due to the insufficiency of the description of the MV state 
by the MF theory. 

Here, we give the following rigorous remarks on the valence transition: 
For general filling of $1/2\le{n}\le1$, the maximum value of 
the jump of $n_{\rm f}$, which is realized for large values of $|\varepsilon_{\rm f}|$ 
and $U_{\rm fc}$, is given by $\Delta n_{\rm f}|_{\rm max}=2n-1$. 
This is
in sharp contrast to the MF result, that always predicts 
the $\rm 4f^1 (Ce^{3+})$ to $\rm 4f^0 (Ce^{4+})$ transition; i.e., 
$\Delta n_{\rm f}|_{\rm max}=1$.
The slope of the FOVT line for large values of $|\varepsilon_{\rm f}|$ 
and $U_{\rm fc}$ 
is expressed by 
$\delta U_{\rm fc}/\delta \varepsilon_{\rm f}=-1$, 
irrespective of total filling $n$~\cite{Watanabe}. 
It is noted that these results hold even 
in the square- and cubic-lattice systems. 

We also point out that 
the reason why the FOVT is difficult to realize in the Ce compounds 
can be understood by the smallness of $U_{\rm fc}$: 
In Ce metal, $U_{\rm fc}$ is large because of the on-site 4f-5d interaction, 
whereas in Ce compounds $U_{\rm fc}$ is considered to be smaller 
even if the coordination number between Ce and surrounding atoms is taken 
into account, 
since $U_{\rm fc}$ corresponds to the intersite 4f-(d, or p) interaction.

We have confirmed that at both sides of the FOVT, 
the peak structure appears at $q=\pi/4$ 
in the Fourier component of $\langle S_{i}^{{\rm a}z} \rangle$ for a=f and c 
when we apply a small magnetic field as a perturbation at the edge sites. 
The above period of the Friedel oscillations corresponds to 
$2k_{\rm F}^{\rm large}=\pi/4$, 
since $k_{\rm F}^{\rm large}=(1+3/4)\pi/2=7\pi/8$. 
The large Fermi surfaces detected in the Kondo and MV states are consistent with the 
existence of the QCP, 
since by detouring around the QCP, the Kondo and MV states can be adiabatically connected, 
with Luttinger's sum rule satisfied. 

In Fig.~\ref{fig:PD}, 
the solid line with solid triangles represents the FOVT line 
determined by the slave boson MF theory. 
It is evident that 
the QCP (solid circle) identified by the DMRG 
is shifted from the QCP (open diamond) by the MF to the 
large-$|\varepsilon_{\rm f}|$ and large-$U_{\rm fc}$ direction. 
We also find no evidence of the phase separation 
at the FOVT, such as negative compressibility, and an inhomogeneous charge 
distribution in the DMRG results for the ground state. 
This result is due to the fact that the order parameter 
$n_{\rm f}$ is not a conserving quantity, which 
is in sharp contrast to the MF theory where the FOVT 
is accompanied by the phase separation~\cite{Watanabe}. 
These results indicate that 
the phase separation is destabilized by quantum fluctuation and electron correlation, 
and as a result the crossover regime from the Kondo to MV states is strongly stabilized. 
In connection to the experiments of the Ce compounds such as $\rm CeCu_2Ge_2$, 
this sharp crossover regime with strong fluctuation is most interesting, 
since the superconductivity is observed in this regime.

\begin{figure}[t]
\begin{center}
\includegraphics[width=8cm]{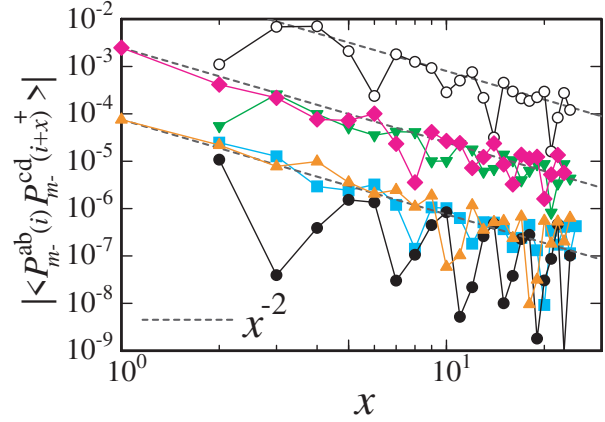}
\end{center}
\caption{
Singlet pairing correlation functions 
$|\langle P_{m-}^{\rm ab}(i){P_{m-}^{\rm cd}}^{\dagger}(i+x) \rangle|$ 
with $P_{m-}^{\rm ab}(i)=(a_{i\uparrow}b_{i+m\downarrow}-
a_{i\downarrow}b_{i+m\uparrow})/\sqrt{2}$ (see text) 
for $\varepsilon_{\rm f}=-3.9$ and $U_{\rm fc}=5.0$ 
at $t=1$, $V=0.1$, $U=100$ and $n=7/8$: 
$|\langle P_{1-}^{\rm cc}{P_{1-}^{\rm cc}}^{\dagger}\rangle|$ 
(black open circle), 
$|\langle P_{1-}^{\rm ff}{P_{1-}^{\rm ff}}^{\dagger}\rangle|$ 
(black solid circle), 
$|\langle P_{0-}^{\rm cf}{P_{1-}^{\rm cc}}^{\dagger}\rangle|$ 
(pink diamond), 
$|\langle P_{1-}^{\rm ff}{P_{1-}^{\rm cc}}^{\dagger}\rangle|$ 
(green triangle), 
$|\langle P_{0-}^{\rm cf}{P_{0-}^{\rm cf}}^{\dagger}\rangle|$ 
(orange triangle) and 
$|\langle P_{1-}^{\rm ff}{P_{0-}^{\rm cf}}^{\dagger}\rangle|$ 
(blue square).
Here, $i=N/2$ is set for $m=0$ and $i=N/2-1$ for $m=1$. 
The dashed line $\sim x^{-2}$ is drawn for comparison. 
}
\label{fig:super}
\end{figure}

To clarify this point, we have calculated the superconducting correlation functions
$\langle P_{m\pm}^{\rm ab}(i){P_{m\pm}^{\rm cd}}^{\dagger}(i+x) \rangle$ 
with $P_{m\pm}^{\rm ab}(i)=(a_{i\uparrow}b_{i+m\downarrow}\pm
a_{i\downarrow}b_{i+m\uparrow})/\sqrt{2}$ for the singlet $(-)$ and triplet $(+)$ 
pairing. 
Here, $a$, $b$, $c$ and $d$  denote the f and/or conduction electron operators and 
we have calculated the nearest-neighbor (NN) pairing $(m=1)$ and 
also the onsite pairing $(m=0)$. 
To avoid the effect of the open boundary, we here fix the position of the operator 
at the central site, i.e., $i=N/2$ for the onsite pairing and 
$i=N/2-1$ for the NN pairing. 
Figure~\ref{fig:super} shows the superconducting correlations 
for $\varepsilon_{\rm f}=-3.9$ at $U_{\rm fc}=5.0$, which is close to 
the sharp valence change (see Fig.~\ref{fig:nf_Ef}(b)). 
A remarkable point here is that the power of the decay of the correlations 
is smaller than 2 as clearly seen, for example, in the data of 
$|\langle P_{0-}^{\rm cf}{P_{1-}^{\rm cc}}^{\dagger} \rangle|$ and 
$|\langle P_{0-}^{\rm cf}{P_{0-}^{\rm cf}}^{\dagger} \rangle|$ 
with dashed lines indicating $x^{-2}$. 
We have confirmed that all correlations of 
$|\langle P_{m-}^{\rm ab}(i){P_{m-}^{\rm cd}}^{\dagger}(i+x) \rangle|$ 
show nearly the same power which is smaller than 2. 
On the other hand, for $\varepsilon_{\rm f}=-4.2$, which is in the Kondo regime, 
we find that the power of the decay of the pairing correlation is larger than 2. 
It is also found that the pairing correlation decays rapidly as $\sim x^{-5}$ for 
$\varepsilon_{\rm f}=-3.71$, where the sharp valence crossover occurs. 
We confirmed that these results are not changed by the calculations 
at several system sizes. 

For $U=U_{\rm fc}=0$, the f level and conduction band are hybridized and 
$\mu$ is located at the lower hybridized band with $k_{\rm F}^{\rm large}$. 
We confirmed that the charge and spin gaps are closed 
with $k_{\rm F}^{\rm large}$ 
in the bulk limit  in both the Kondo, MV and crossover regimes. 
Although the justification requires further careful studies, we assume the system 
being described by the single-band Tomonaga-Luttinger (TL) liquid in this Letter. 
For the single-component TL liquid without the spin and charge gaps, 
the singlet pairing correlation decays as $x^{-1-1/K_{\rho}}$~\cite{Schultz}. 
Since the charge and spin correlations decay as $x^{-1-K_{\rho}}$~\cite{Schultz}, 
the above result implies that $K_{\rho}>1$, namely,  
that the superconducting correlation becomes dominant. 

In order to estimate the TL parameter $K_{\rho}$ systematically, 
we made a least-squares fit for the charge, spin and superconducting correlations 
defined by 
$\langle \bar{O}_{i} \bar{O}^{\dagger}_{j} \rangle$ for 
$\bar{O_i}\equiv{O_i}-\langle{O_i}\rangle$ 
and $O_i=n^{\rm c}_i$, $S^{{\rm c}z}_i$ and 
$P^{\rm cc}_{m-}(i)$, respectively, 
since the conduction electrons have a large amplitude in the pairing correlation 
as seen in Fig.~\ref{fig:super}, which seems to give a reliable estimate. 
By assuming the asymptotic forms of the correlation functions of the 
single-component TL liquid~\cite{Schultz}, 
we obtained three $K_{\rho}$'s from the singlet-superconducting (SS), 
the charge-density-wave (CDW) and the spin-density-wave (SDW) correlation functions. 
Then, we take the middle point between their maximum and minimum values as a mean value, 
with the error defined by the difference 
between the middle point and the maximum (minimum) value.

The $K_{\rho}$ values evaluated in this manner are shown in Fig.~\ref{fig:nf_Ef} 
as the solid diamonds. 
Here, for $\varepsilon_{\rm f}{\ge}-3.75$ at $U_{\rm fc}=5.0$ and 
for $\varepsilon_{\rm f}{\ge}-5.6$ at $U_{\rm fc}=7.0$, $K_{\rho}$ is estimated from 
4$k_{\rm F}$CDW and SS correlations. 
For $U_{\rm fc}=5.0$, $K_{\rho}$ appears to exceed 1 
at around $\varepsilon_{\rm f}=-3.9$ in Fig.~\ref{fig:nf_Ef}(b), 
suggesting that the superconducting correlation is dominant, 
which is consistent with the result presented in Fig.~\ref{fig:super}. 
For $U_{\rm fc}=7.0$, $K_{\rho}$ jumps at the FOVT 
as in Fig.~\ref{fig:nf_Ef}(a). 
For $U_{\rm fc}=0$, $K_{\rho}$ estimated from the SS correlation 
remains smaller than 1, indicating that the pairing correlation 
does not develop as in Fig.~\ref{fig:nf_Ef}(c). 
These results show that the superconducting correlation is enhanced 
in the Kondo regime near the sharp valence crossover, 
which is close to the QCP of the valence transition. 

\begin{figure}[t]
\begin{center}
\includegraphics[width=8cm]{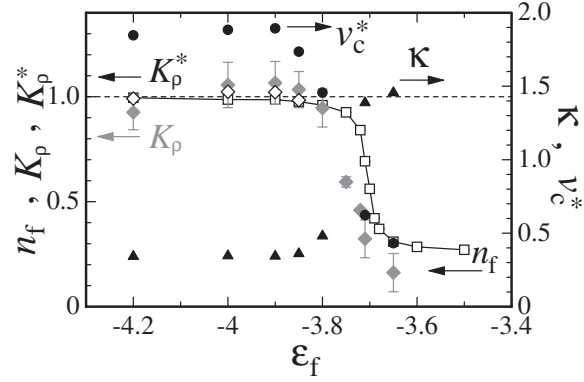}
\end{center}
\caption{
Compressibility (solid triangle) obtained by 
$\kappa=\lim_{N\to\infty}2/(N\Delta_{\rm c})$ and 
charge velocity (solid circle) obtained by 
$v_{\rm c}^{*}=\lim_{N\to\infty}\Delta_{\rm c}/\Delta q$ 
for $t=1$, $V=0.1$, $U=100$ and $U_{\rm fc}=5.0$ at $n=7/8$. 
The $K_{\rho}$ (shaded diamond) obtained by the least-squares fit of 
correlation functions, $K_{\rho}^{*}=\pi v^{*}_{\rm c}\kappa/2$ (open diamond), 
and $n_{\rm f}$ (open square) are shown. 
}
\label{fig:nK}
\end{figure}

To get further insight into the nonmonotonic behavior of $K_{\rho}$ around 
$\varepsilon_{\rm f}=-3.9$ at $U_{\rm fc}=5.0$, 
we calculated the charge compressibility 
$\kappa \equiv \partial (2n)/\partial \mu$, 
since the TL parameter is expressed by the charge velocity $v_{\rm c}$ and $\kappa$ 
as $K_{\rho}=\pi v_{\rm c}\kappa/2$. 
In Fig.~\ref{fig:nK}, we plot 
the compressibility extrapolated to the $N{\to}{\infty}$ limit in the form of 
$\kappa=2/(N\Delta_{\rm c})$ with the charge gap 
$\Delta_{\rm c}\equiv[E(2nN+2,0)+E(2nN-2,0)-2E(2nN,0)]/2$, 
where $E(N_{\rm e},S)$ is the ground-state energy for the electron number $N_{\rm e}$ 
and total spin $S$. 
As $\varepsilon_{\rm f}$ increases, $\kappa$ increases, 
since total charge fluctuation becomes large. 
We see that $\kappa$ does not show any diverging enhancement at the sharp valence 
crossover, $\varepsilon_{\rm f}{\sim}-3.71$, in contrast to the enhanced $\chi_{\rm f}$. 
This is due to the cancellation of the leading terms between 
$\langle \bar{n}_{i}^{\rm a}\bar{n}_{j}^{\rm a} \rangle$ for a=f,c 
and 
$\langle \bar{n}_{i}^{\rm f}\bar{n}_{j}^{\rm c} \rangle$ 
in 
$\langle (\bar{n}_{i}^{\rm f}+\bar{n}_{i}^{\rm c})
(\bar{n}_{j}^{\rm f}+\bar{n}_{j}^{\rm c}) \rangle$, 
since the diagonal and off-diagonal charge correlations have the opposite signs 
for $U_{\rm fc}>0$. 

A remarkable result here is that the compressibility stays constant 
around $\varepsilon_{\rm f}=-3.9$, where $K_{\rho}$ 
has a maximum. 
This indicates that the enhancement of $K_{\rho}$ is caused by the enhancement 
of the charge velocity, but not by the compressibility. 
This is quite different from the ordinary models, which exhibits $K_{\rho}>1$~\cite{ogata}.
In order to get further insight into this finding, we calculated the charge velocity 
defined by $v_{\rm c}^{*}\equiv \lim_{N \to \infty}\Delta_{\rm c}/\Delta q$ 
with $\Delta q=\pi/(N+1)$ 
instead of $v_{\rm c}$~\cite{Kc}, since the numerical accuracy is not enough for 
the second-lowest eigenvalue in the Lanczos diagonalization. 
We see that the maximum of $v_{\rm c}^{*}$ appears around $\varepsilon_{\rm f}=-3.9$ 
and this suggests that the enhancement of $v_{\rm c}$ 
is the origin of the enhancement of $K_{\rho}$. 
In Fig.~\ref{fig:nK}, we show $K_{\rho}^{*}=\pi \kappa v_{\rm c}^{*}/2$ 
for $\varepsilon_{\rm f}{\le}-3.85$ by open diamonds and 
note that $K_{\rho}^{*}$ 
coincides with the $K_{\rho}$ obtained by the least-squares fit 
of the correlation functions within the error bars. 
Since $K_{\rho}^{*}$ is obtained by $v_{\rm c}^{*}$ and $\kappa$ after the 
extrapolation to $N\to\infty$, 
this suggests that $K_{\rho}$ is evaluated properly within the finite sizes 
calculated in this study. 
A more accurate determination of $K_{\rho}$ in the bulk limit 
is left for future study.

The increase in $v_{\rm c}^{*}$ toward $\varepsilon_{\rm f}{\simeq}-3.9$ in the Kondo regime 
is considered to be due to the decrease in the density of states 
at the Fermi level, i.e., the broadening of the Kondo peak. 
As $\varepsilon_{\rm f}$ further increases, $v_{\rm c}^{*}$ is suppressed, 
since $U_{\rm fc}$ interrupts the coherent motion of the 
electrons in the MV state.  
This is the reason why $v_{\rm c}^{*}$ has a maximum near the sharp valence increase. 
The enhanced $v_{\rm c}^{*}$ implies the enlargement of the effective band width 
and thus the coherent motion of electrons with enhanced valence fluctuation 
is considered to be the origin of the development of the pairing correlation. 
The enhancement of $K_{\rho}$ around $\varepsilon_{\rm f}=-3.9$ 
also agrees with the following naive expectation: 
In the deep Kondo regime, the SDW correlation is dominant by the 
RKKY interaction 
and at the sharp valence crossover point the CDW correlation becomes dominant 
with enhanced $\chi_{\rm f}$. 
Between the two regimes, the superconducting correlation can be dominant, 
since the regime midway between the Kondo state and the sharp valence crossover 
is unfavorable for both the SDW and CDW correlations.

To summarize, we have clarified the novel nature of 
the electronic states near the QCP of the valence transition. 
Since valence fluctuations are basically ascribed to atomic origin, 
we believe that 
the basic nature of the valence transition and effect of its fluctuation 
revealed in this study is universal in two- and three-dimensional systems as well. 

One of the authors (S. W.) thanks H. Harima and 
S. Kawasaki for fruitful discussions. 
This work is supported in part by the Creative Research Fund (No. 15G0213) from 
the Japan Society for the Promotion of Science.



\begin{thebibliography}{9}
\bibitem{Kc} When the charge velocity $v_{\rm c}$ is defined by 
$v_{\rm c}\equiv \lim_{{\Delta}q \to 0} \Delta_{\rm c}^{*}/\Delta q$ 
for $\Delta_{\rm c}^{*}\equiv E_{1}(2nN,0)-E(2nN,0)$ 
with $E_{1}$ being the first-excited state, 
$K_{\rho}=\pi v_{\rm c}\kappa/2$ holds. 
If the symmetric condition $\varepsilon_{\rm f}+U/2=0$ and $U_{\rm fc}=0$ 
are satisfied 
in eq.~(\ref{eq:PAM}), 
$\Delta_{\rm c}=\Delta_{\rm c}^{*}$; i.e., $v_{\rm c}=v_{\rm c}^{*}$ 
holds by charge SU(2) symmetry. 
It is expected that $v_{\rm c} \sim v_{\rm c}^{*}$ holds for eq.~(\ref{eq:PAM}) 
in the Kondo regime.
\end{thebibliography}
\end{document}